\RequirePackage{fix-cm}
\documentclass[smallextended]{svjour3}       % onecolumn (second format)
\smartqed  % flush right qed marks, e.g. at end of proof
\usepackage{graphicx}
\begin{document}

\title{Model of a Solar System in the Conservative Geometry}

%\titlerunning{Short form of title}        % if too long for running head

\author{Edward Lee Green }

%\authorrunning{Short form of author list} % if too long for running head

\institute{E. Green \at
              University of North Georgia, Dahlonega, GA 30597 \\
              Tel.: +706-864-1809\\
              Fax: +706-864-1678\\
              \email{egreen@ung.edu}           %  \\
%             \emph{Present address:} of F. Author  %  if needed
}

\date{Received: date / Accepted: date}
% The correct dates will be entered by the editor

\maketitle

\begin{abstract}
Pandres has shown that an enlargement of the covariance group to the group of conservative transformations leads to a richer geometry than that of general relativity.  Using orthonormal tetrads as field variables, the fundamental geometric object is the curvature vector denoted by $C_\mu$.  From an appropriate scalar Lagrangian field equations for both free-field and the field with sources have been developed.  We first review models which use a free-field solution to model the Solar System and why these results are unacceptable.  We also show that the standard Schwarzschild metric is also unacceptable in our theory.  Finally we show that there are solutions which involve sources which agree with general relativity PPN parameters and thus approximate the Schwarzschild solution.  The main difference is that the Einstein tensor is not identically zero but includes small values for the density, radial pressure and tangential pressure.  Higher precision experiments should be able to determine the validity of these models.  These results add further confirmation that the theory developed by Pandres is the fundamental theory of physics.
%\keywords{First keyword \and Second keyword \and More}
 \PACS{98.80.-k \and 04.20.Ha \and 04.20.Jb \and  12.10.-g}
% \subclass{MSC code1 \and MSC code2 \and more}
\end{abstract}

\section{Introduction}

We assume a 4-dimensional space with the orthonormal tetrad $h^{i}_{\,\,\mu}$ and internal tetrad $L^i_{I}$ as field variables.   (Please refer to earlier papers by Pandres and Green for a thorough discussion of the physical motivation and for additional examples of how to interpret the theory \cite{Pand81,Pand84,Pand09,Green09,PG03,Green11}.) In the present paper, which is concerned with gravity and cosmology, the internal tetrad may be chosen to be constants.   The metric is defined by $g_{\mu\nu}=\eta_{ij}\, h^i_{\,\,\mu}h^j_{\,\,\nu}$ where $\eta_{ij} =
diag\bigl\{-1,1,1,1\bigr\}$.   (Note: our sign conventions are the same as Misner, Thorne and Wheeler \cite{MTW}.)   The following condition (1) defines the conservation group, a group of coordinate transformations that include the group of diffeomorphisms as a proper subgroup \cite{Pand81}:
\begin{equation}x^\nu_{\;\; ,\overline{\alpha}}\bigl(x^{\overline{\alpha}}_{\;\; ,\nu ,\mu} -
x^{\overline{\alpha}}_{\;\; ,\mu , \nu} \bigr)\; = \; 0 \qquad .  \label{conserv} \end{equation} This group preserves the wave equation and other conservation laws \cite{Pand84}, which are of the form $V^\alpha_{\,\, ;\alpha}=0\,$.

The geometry determined by (\ref{conserv}) is more general than a manifold \cite{Pand81,Pand84,Pand09,Green09,PG03}.  We regard a conservative, non-diffeomorphic transformation as a mapping between two manifolds \cite{Green11}. The geometrical content of space-time is determined by the curvature vector:
\begin{equation}  C_\alpha \equiv \; h_i^{\,\,\nu}\bigl(h^i_{\,\, \alpha ,\nu}-h^i_{\,\, \nu ,\alpha}
\bigr) \; = \; \gamma^\mu_{\;\;\alpha\mu} \label{curv} \end{equation}
where the Ricci rotation coefficient is given
by $\gamma^i_{\;\;\mu\nu}=h^i_{\;\mu ;\nu}$ \cite{Pand81,Pand84,Pand09}. The curvature vector, $C_\alpha$, is covariant under transformations from $x^\mu$ to $x^{\overline{\mu}}$ if and only if the transformation is conservative, i.e., it satisfies (\ref{conserv}).

When sources are present, the Lagrangian is of the form
\begin{equation} {\cal L}= {\cal L}_{\rm f}+{\cal L}_{\rm s} =  \int \biggl( \frac1{16\pi}C^\alpha C_\alpha + L_s \biggr) \, h \; d^4 x  \label{e10} \end{equation}  where $L_s$ is the appropriate source Lagrangian density function.  Using the Ricci rotation coefficients, one finds that
$C^\alpha C_\alpha = R + \gamma^{\alpha \beta \nu} \gamma_{\alpha \nu \beta}-2C^\alpha_{\; ;\alpha}
\; $
where $R$ is the usual Ricci scalar curvature.   Thus, the free-field part contains additional terms which may correspond to other forces or dark matter \cite{Pand09,Green09,Green11}.  Variation of (\ref{e10}) with respect to the tetrad results in (see \cite{Wein72})
$$\int \Biggl[\frac{1}{16\pi}\biggl(2C_{(\mu ;\nu)}- 2C_\alpha \gamma^\alpha_{\; (\mu\nu)} - g_{\mu\nu}C^\alpha C_\alpha - 2g_{\mu\nu}C^\alpha_{\; ;\alpha} \biggr) - (T_{\rm s})_{\mu\nu}\Biggr] h\, h^{i\nu} \delta h_i^{\;\mu} \, d^4 x = 0$$
Thus the source stress-energy tensor is
\begin{equation}  8\pi (T_{\rm s})_{\mu\nu} \; = \;   C_{(\mu ;\nu)}- C_\alpha \gamma^\alpha_{\; (\mu\nu)} -\frac12 g_{\mu\nu}C^\alpha C_\alpha - g_{\mu\nu}C^\alpha_{\; ;\alpha}  \label{e11} \end{equation}
and we also find
\begin{equation} G_{\mu\nu}=  \biggl( \gamma^{\;\;\alpha}_{(\mu \;\; \nu ) ;\alpha}+\gamma^\alpha_{\;\;\sigma
 (\nu} \gamma^\sigma_{\;\; \mu ) \alpha} + \frac12 g_{\mu\nu}\gamma^{\alpha \beta
\sigma} \gamma_{\alpha \sigma \beta} \biggr) + 8\pi (T_{\rm s})_{\mu\nu}  \quad   \label{e12} \end{equation}
The first term in the parentheses represents the free-field part of the total stress energy tensor which is the dark matter/dark energy part.  The second term given by (\ref{e11}) transforms covariantly under the conservation group (\ref{conserv}).

\section{Spherically symmetric solutions.}

In this section we develop the formulae for the curvature vector, metric and total stress-energy tensor.  In spherical coordinates, an arbitrary spherically symmetric tetrad may be expressed by
\begin{equation}h^i_{\;\; \mu} =  \left[ \begin{array}{cccc}
\; e^{\Phi(r)} & 0 & 0 & 0 \\
0 &\; e^{\Lambda(r)}\sin \theta \cos \phi \; & \; r\cos\theta\cos\phi \; & \; -r\sin\theta\sin\phi \; \\
0 & e^{\Lambda(r)}\sin\theta\sin\phi & r\cos\theta\sin\phi    & \; \;r\sin\theta\cos\phi \\
0 & e^{\Lambda(r)}\cos\theta\qquad & -r\sin\theta\qquad & 0
\end{array}
\right]
\label{e31} \end{equation}
where the upper index refers to the row. The curvature vector for this tetrad field is given by
\begin{equation} C_\mu = \frac{e^{\Lambda}}r \biggl[ \; 0,  \; 2 - e^{-\Lambda}\bigl(r\Phi^\prime + 2\bigr) , \; 0 , \;0  \biggr]  \label{e32} \end{equation}
where components are in the order $[t,r,\theta,\phi]$ and the prime denotes the derivative with respect to $r$. The tetrad (\ref{e31}) leads to the metric
\begin{equation}ds^2= -e^{2\Phi(r)} dt^2 +  e^{2\Lambda(r)}dr^2+r^2d\theta^2 +r^2\sin^2\theta d\phi^2 \quad . \label{e33} \end{equation}
When $(r\Phi^\prime + 2)=2e^{\Lambda}$, then $C_\mu$ in equation (\ref{e32}) is identically zero and hence $(T_{\rm s})_{\mu\nu}  $ is identically zero leading to a free-field solution.

The metric (\ref{e33}) leads to a diagonal Einstein tensor with nonzero elements:
\begin{equation}G^t_{\, t}\equiv -8\pi \rho  \; = \, \frac1{r^2}\bigl(-2re^{-2\Lambda}\Lambda^\prime + e^{-2\Lambda} - 1 \bigr) \;  = -\frac2{r^2}\frac{d}{dr}\biggl[\frac12 r \Bigl(1  - e^{-2\Lambda}\Bigr) \biggr]  \label{e34} \end{equation}
\begin{equation}G^r_{\, r} \equiv \; 8\pi p_r \; = \; \frac1{r^2}\Bigl( 2re^{-2\Lambda}\Phi^\prime + e^{-2\Lambda}-1\Bigl)  \label{e35} \end{equation}
\begin{equation} G^\theta_{\, \theta}\, =\; G^\phi_{\, \phi} \, \equiv \; 8\pi p_T \, = \; \frac{e^{-2\Lambda}}{r}\biggl(r\Phi^{\prime\prime}+r(\Phi^\prime)^2 - r\Phi^\prime \Lambda^\prime + \Phi^\prime - \Lambda^\prime \biggr)  \label{e36} \end{equation}

We note that $G^t_{\, t}= -8\pi \rho$ depends only on $\Lambda(r)$.  Depending on whether $C_\mu$ is zero via (\ref{e32}) we will have a free-field or a field with sources.    The source term of the Lagrangian (\ref{e10}) that we will use is $L_s=\rho_s(r)$, where $\rho_s(r)$ is the density of the source as a function of $r$.  Previous results suggest the interpretation that the density of the source is given by \cite{Green11}:
\begin{equation}
  8\pi \rho_s = \, \frac12 C^\mu C_\mu \,   \label{rhos}
\end{equation}
We assume that for realistic densities,  $\rho_s < \rho$ and we define $\rho_{dm} \equiv \rho - \rho_s$ to represent the density of dark matter.

\subsection{Particle motion in the spherically symmetric case.}

We exhibit here the general formulae governing such motion in a spherically symmetric solution of our field equations.  The equations of motion for a particle are based on adding an appropriate term to the Lagrangian.  Using an approximate delta function, $\delta_\epsilon^4$, and following \cite{dFC90} we add a term
\begin{equation}  L_{p} = \rho_{p}(x)  = \mu \int \delta_\epsilon^4(x-\gamma(s))(-u^\mu u_\mu)^{\frac12} ds   \label{e67} \end{equation}
to the Lagrangian of (\ref{e10}). Note: the space-like volume is represented by $\epsilon$ which is assumed to be small.  The mass of the particle will be denoted by $\mu$.  The path of the particle is represented by $\gamma(s)$ and its velocity is $u^\alpha=\frac{dx^\alpha}{d\tau}$.  We will use the "dot" notation for the components of $u^\alpha$, i.e. $u^\alpha = \langle \dot{t},\dot{r},\dot{\theta},\dot{\phi}\rangle$.  The condition  $T^{\beta \alpha}_{\; \;\;\; ; \beta }=0$ leads to (see \cite{dFC90} and \cite{Green11})
\begin{equation}  \; \frac{\mu}{\sqrt{-u^\nu u_\nu}} \, u^\beta u^\alpha_{\; ;\beta} \, = \; \delta^\alpha_1 F_p   \label{e68} \end{equation}
with $F_p \equiv \epsilon \, e^{-2\Lambda} \bigl(-p_R^{\; \prime} + \frac2r \, \bigl(p_T-p_R\bigr)\,\bigr) $.  When $\alpha\neq 1$, (\ref{e68}) implies  $u^\beta u^\alpha_{\; ;\beta} = 0$ which is the usual geodesic equation.

We will only be concerned with orbits and motions that can be confined to a surface which we henceforth assume to be the $\theta = \frac{\pi}2$ surface.  This is consistent with $\theta$ component of (\ref{e68}) which implies that
$   \ddot{\theta} + \frac2r \, \dot{r}\, \dot{\theta} - (\sin\theta \cos\theta) \dot{\phi}^2 =0  \,$.
The $\phi$ component of (\ref{e68}) using metric (\ref{e33}) implies that
\begin{equation}   \ddot{\phi} + \frac2r \, \dot{r} \, \dot{\phi} \, = \, 0 \quad  \qquad  \Rightarrow \quad  \qquad  \dot{\phi} \; = \; \frac{L}{r^2} \label{phidot} \end{equation}
where $L$ is a constant interpreted as the conserved angular momentum.
From (\ref{e68}) we also find that the $\, t\,$ component is
\begin{equation}  \ddot{t} \, + \, 2 \dot{\Phi} \, \dot{t} \, = \, 0 \quad \qquad \Rightarrow \quad \qquad \dot{t} \, = \, E \, e^{-2\Phi} \label{tdot} \end{equation}
where $\Phi(r)$ is the function appearing in $h^0_{\, 0} \, $ (see \ref{e31}) and $E$ is a constant (energy) of the motion.
For the metric (\ref{e33}) with $\theta \equiv \frac{\pi}2$, the $r$ component of the particle motion is determined by
\begin{equation}  \ddot{r} \, + \, e^{2\Phi-2\Lambda}\Phi^\prime (r) \, \dot{t}^2 \, + \, \Lambda^\prime (r) \, \dot{r}^2 \, - \, r \, e^{-2\Lambda}  \dot{\phi}^2 \, = \; \frac1{\mu} F_p \label{rddot}
\end{equation}
Using (\ref{phidot}), (\ref{tdot}), (\ref{e35}) and (\ref{e36}) along with the normalization $\dot{t}=e^{-2\Phi}$ (i.e., we choose $E=1$ in (\ref{tdot}) - see \cite{Wein72}, page 186), we find that
\begin{equation}  \ddot{r} \, +  \Lambda^\prime \dot{r}^2  + e^{-2\Phi-2\Lambda} \Phi^\prime - \frac{e^{-2\Lambda}}{r^3} L^2 \, = \, \frac{2\, \epsilon\, e^{-4\Lambda}\Phi^\prime (\Lambda^\prime + \Phi^\prime)\,}{\mu\, r}  \label{rddot2} \end{equation}

\section{Free-field models for the Solar System.}

We first consider the free-field case where $C^\mu = 0$.  This implies that $C^\mu C_\mu = 0$ and hence the density of the source, $\rho_s$ is identically zero.  Let $w(r)=\frac{r}{2}\Bigl(1-e^{-2\Lambda}\Bigr)$ and assume that ${\lim_{r\to\infty} w(r) = W}$, a constant which is related to the asymptotic value of the mass of the Sun.  In this case, using (\ref{e32}),
\begin{equation}
  \Phi^\prime(r) = \frac2r\biggl( e^\Lambda - 1 \biggr) \; = \;  \frac2r\biggl(1-\frac{2w(r)}r \biggr)^{-\frac12} - \frac2r
\end{equation}
Asymptotically, we have
\begin{equation}
  \Phi^\prime(r) \; \approx \; \frac{2W}{r^2} \quad \Rightarrow  \quad  e^{2\Phi(r)} \approx   1- \frac{4W}r
\end{equation}
To match with the weak field solution of general relativity, we thus conclude that $W=\frac12 M$, where $M$ is the total mass-energy of the Sun.  Thus we see that in the free-field models,  $G^t_{\, t}$ represents $\frac12$ of the density of mass-energy (nevertheless, we shall continue to denote $G^t_{\, t}$ by $-8\pi \rho$ in the free-field case).

In \cite{Green11}, we investigated this case in-depth and found several positive results.  The tetrad solutions that were investigated were determined by $w(r)=\frac12 M$ (Case 1) and $w(r)=\frac{M}2 - \frac{M^2}{8r}$ (Case 2).  In both cases we found that the Einstein tensor was nonzero (see (\ref{e34}) - (\ref{e36})). In Case 1, asymptotically, we had $8\pi\rho = 0$,  $8\pi p_r \approx \frac{M}{r^3}$ and $8\pi p_T \approx -\frac{M}{2r^3}$.  In Case 2, we had the same asymptotically for $p_r$ and $p_T$, but we had $8\pi \rho = \frac{M^2}{4r^4}$.

We found asymptotic agreement with the gravitational redshift and with Kepler's Law for both Case 1 and Case 2.  Additionally, we were able to explain that the Pioneer anomaly was produced by the nonzero and nonequal radial and tangential pressures.  About the same time, another explanation of the Pioneer anomaly was given in terms thermal corrections \cite{Turyshev}.

However, discrepancies were found with the precession of perihelia. Specifically, the precession of perihelia calculation for Case 1 produced a value which was $\frac58$ of the standard value and the calculation for Case 2 produced a value which was $\frac7{12}$ of the standard value.  Arguments were made to justify the discrepancy in Case 2, since the nonzero density could contribute to the precession.

We also found discrepancies in the light deflection and time delay predictions.  For both the light deflection and time delay calculations our theory yielded values which were 75\% of the standard value (for both Case 1 and Case 2). It was argued that the nonzero density of the inertial mass (given by $\rho + p_r$) along with reasonable refraction index values could explain the discrepancy.  It is difficult to imagine why the numbers would agree with the standard result as observed in solar system experiments, so we reject the free-field model due to a lack of naturalness.

\section{Models for the Solar System which include sources.}

When setting up a model with our theory, one should keep in mind that the there is a great deal of freedom due to the larger covariance group.  This means that additional conditions may be needed to obtain a model that classically represents a given physical situation.   One may view solutions that do not meet these conditions as being non-physical or as non-classical.  The additional conditions given below will enable us to obtain a reasonable model of the Solar System.

Recall that the scalar formed from the curvature vector (\ref{curv}) by $C^\mu C_\mu$ exactly determines the density of mass-energy of the sources according to $8\pi \rho_s = \, \frac12 C^\mu C_\mu \, $ \cite{Green11}.  We note that $C^\mu$ is a vector under the group of conservative transformations and thus $C^\mu C_\mu$ is an invariant scalar.   From (\ref{e32}) and (\ref{e33}) we find that $C^\mu C_\mu = \, e^{-2\Lambda}\biggl(\, \frac{\, 2e^\Lambda}{r}-\Phi^\prime - \frac2r \;\biggr)^2$.

For convenience in our model development, we define $\beta(r) = \sqrt{C^\mu C_\mu}$.  Hence
\begin{equation}
   \Phi^\prime =  \frac{\, 2e^\Lambda}{r} - \frac2r \, - \, \beta(r) e^\Lambda \label{Phiprime}
\end{equation}
As a strategy for model development, we first will propose a function for $\Lambda(r)$, or equivalently $e^{2\Lambda}$.   We will require that, in contrast to the free-field case, $G^t_{\, t}$ corresponds to $-8\pi \rho$, the usual density of mass-energy (\ref{e34}).  (See condition C1) below.) This implies that
\begin{equation}
  m(r) = \frac12 r (1-e^{-2\Lambda} ) \label{mass}
\end{equation}
The next factor (see condition C2) below) to consider is whether the corresponding metric asymptotically agrees with the weak-field metric, i.e., $g_{tt} \approx - (1 - \frac{2M}r )$.  The third factor to consider (see condition C3) below) is whether the density of dark matter is nonnegative, i.e., $\rho - \rho_s \geq 0$.

Obviously the standard Schwarzschild metric is the first example to consider.  Accordingly we choose $e^{2\Lambda} = \frac1{1-\frac{2M}r}$ and $e^{2\Phi} = 1-\frac{2M}r$.  From (\ref{Phiprime}) we then find $\beta = \frac{(1-\sqrt{1-\frac{2M}r})(1-3\sqrt{1-\frac{2M}r})}{2r\sqrt{1-\frac{2M}r}}$  and hence
\begin{equation}
  C^\mu C_\mu \; = \; \frac{(1-\sqrt{1-\frac{2M}r})^2(1-3\sqrt{1-\frac{2M}r})^2}{4r^2(1-\frac{2M}r)} \approx  \frac{M^2(1-\frac{M}r )}{r^4} \label{schwarzCmuCmu}
\end{equation}
where the approximation is for $r>>M$.

This solution is not acceptable however, because $G^{t}_{\, t} = 0$ and thus $\rho \equiv 0$ and yet $\rho_s \approx \frac{M^2(1-\frac{M}r)}{16\pi r^4}$.  Since $\rho_s + \rho_{dm} =\rho$, this implies that $\rho_{dm} < 0$.  If we require that the density of mass-energy be positive in the case of the solar system (no extreme energies, such as extreme gravity near a black hole) then we must abandon this solution as being physically unreasonable.

The Parameterized-Post Newtonian (PPN) formulism utilizes a parameterized metric which is in the isotropic form.  First we exhibit the general spherical isotropic tetrad (where $\Phi$ represents $\Phi(r)$ and $\Lambda$ represents $\Lambda(r)$):
\begin{equation}h^i_{\;\; \mu} =  \left[ \begin{array}{cccc}
\; e^{\Phi} & 0 & 0 & 0 \\
0 &\; e^{\Lambda}\sin \theta \cos \phi \; & \; r e^{\Lambda}\cos\theta\cos\phi \; & \; -r e^{\Lambda}\sin\theta\sin\phi \; \\
0 & e^{\Lambda}\sin\theta\sin\phi & r e^{\Lambda}\cos\theta\sin\phi    & \; \;r e^{\Lambda}\sin\theta\cos\phi \\
0 & e^{\Lambda}\cos\theta\qquad & -r e^{\Lambda}\sin\theta\qquad & 0
\end{array}
\right]
\label{tetradiso} \end{equation}
where the upper index refers to the row. The curvature vector for this tetrad field is given by
\begin{equation} C_\mu =  \biggl[ \; 0 \; ,  \; - \, \frac{d \Phi}{dr}  - \, 2 \, \frac{d \Lambda}{dr} \;   , \; 0 \; , \;0  \; \biggr]  \label{Cmuiso} \end{equation}
where components are in the order $[t,r,\theta,\phi]$. Hence we have
\begin{equation}
  C^\mu C_\mu \; = \; e^{-2\Lambda} \biggl( \; \frac{d\Phi}{dr} \, + \, 2 \, \frac{d\Lambda}{dr} \; \biggr)^{ 2} \label{CmuCmuiso}
\end{equation}
This tetrad (\ref{tetradiso}) yields a spherically symmetric metric in isotropic coordinates given by
\begin{equation}
  ds^2 = \; - e^{2\Phi} dt^2 + \; e^{2\Lambda}\biggl[ dr^2 + r^2 d\theta^2 + r^2 \sin^2\theta d\phi^2 \biggr] \label{isogenmetric}
\end{equation}
The metric (\ref{isogenmetric}) leads to a diagonal Einstein tensor with nonzero elements:
\begin{equation}
   G^t_{\, t} \equiv  \, \frac{e^{-2\Lambda}}{r} \biggl[ \;   4 \Lambda^\prime + 2 r \Lambda^{\prime\prime} + r (\Lambda^\prime)^2 \, \biggr] \;   \label{Gttiso}
\end{equation}
\begin{equation}G^r_{\, r} \equiv  \; \frac{e^{-2\Lambda}}{r} \biggl[ \; 2\Phi^\prime + 2\Lambda^\prime + 2r\Phi^\prime \Lambda^\prime + r (\Lambda^\prime)^2 \,  \biggr]   \label{Grriso} \end{equation}
\begin{equation} G^\theta_{\, \theta}\, =\; G^\phi_{\, \phi} \, \equiv \; \frac{e^{-2\Lambda}}{r} \biggl[ \;  \Phi^\prime +\Lambda^\prime + r \Phi^{\prime\prime} + r (\Phi^\prime)^2 + r (\Lambda^\prime)^2 \, \biggr]  \label{Gphiphiiso} \end{equation}
where primes indicate derivatives with respect to $r$.

In the PPN system, the value of $e^{2\Phi}$ is parameterized up to second order in $\frac{M}r$, but the value of $e^{2\Lambda}$ is specified only up to first order in $\frac{M}r$.  The unacceptable solution  (\ref{schwarzCmuCmu}) for exact matching with the Schwarzschild metric leads us to take a different approach.  We start with the isotropic form (\ref{isogenmetric}) with
\begin{equation}
  e^{2\Lambda}  \approx  1 + \frac{2M}r + \frac{bM^2}{r^2} \label{eLambdaiso}
\end{equation}
where $b$ is a free parameter.  We also require that our resulting metric (in isotropic coordinates) agrees with the standard external Schwarzschild solution so that there will be agreement with Solar System experiments.  Thus
\begin{equation}
  e^{2\Phi} \approx 1 - \frac{2M}r + \frac{2M^2}{r^2}  \label{ePhiiso}
\end{equation}
We now proceed to find values for $\rho_s$ and $\rho$ up to the terms of order $\frac{M^3}{r^5}$ (note:  $\frac{M}r$ is of order $10^{-7}$ in natural units). We first calculate $\rho_s$ using (\ref{rhos}) and (\ref{CmuCmuiso}) and inserting (\ref{ePhiiso}) and (\ref{eLambdaiso}).  The result is
\begin{equation}
  8\pi \rho_s  \, \approx \; \frac{M^2}{2r^4} \; + \; \frac{(4b-10)M^3}{r^5} \label{rhosPPN}
\end{equation}
We will use a {\it bar} over the radial coordinate ($\bar{r}$) to represent the fact that we are using the standard radial coordinate and use the usual symbol, $r$, for isotropic coordinates.  The conversion formula is simply,
\begin{equation}
  \bar{r} \; = \; r \, e^{\Lambda(r)}   \qquad   \Rightarrow  \qquad  r  \; \approx  \;  \bar{r} \biggl( \; 1 - \frac{M}{\bar{r}} + \frac{(1-b)M^2}{2\bar{r}^2} \; \biggr)  \label{r_rbar}
\end{equation}
Because of (\ref{r_rbar}) the expression for $g_{tt} = -e^{2\Phi}$ depends on $b$:
\begin{equation}
  e^{2\Phi} \approx \; 1 - \frac{2M}{\bar{r}} + \frac{(4-2b)M^3}{\bar{r}^3}
\end{equation}
(notice that the term of order $\frac{M^2}{\bar{r}^2}$ is zero).  From (\ref{rhosPPN}) we find that in the standard radial coordinates,
\begin{equation}
  8 \pi \rho_s \, \approx \; \frac{M^2}{2\bar{r}^4} \, + \, \frac{(2b-3)M^3}{\bar{r}^5}  \label{rhosapprox}
\end{equation}
We now find the value of $\rho$ (total density) for (\ref{isogenmetric}) with conditions (\ref{eLambdaiso}) and (\ref{ePhiiso}).  The result is
\begin{equation}
  8\pi \rho \; \approx \; - \, \frac{(2b-3)M^2}{r^4} + \frac{4(2b-3)M^3}{r^5} \label{rhoPPN}
\end{equation}
After converting to standard coordinates we find
\begin{equation}
  8 \pi \rho \; \approx \; - \, \frac{(2b-3)M^2}{\bar{r}^4}
\end{equation}
where the $\frac{\, M^3}{\bar{r}^5}$ term is zero.

\subsection{Acceptability Conditions.}

For the model to be acceptable, we require that its:

\phantom{D}

\qquad C1)  Einstein tensor component $G^t_t$ match with $\, -\, 8\pi \rho\,$;

\qquad C2)  metric match with the asymptotic (weak field) approximation; and

\qquad C3) (since $8\pi \rho_s = \frac12 C^\mu C_\mu$) value of $C^\mu C_\mu$ must be such that $\, \rho_s < \rho\,$.

\phantom{D}

Conditions C1) and C2) imply that $C_\mu$ is nonzero.  Thus in order that our Solar System models be acceptable, $C^\mu C_\mu$ must be nonzero with $8\pi \rho_s = \frac12 C^\mu C_\mu < -G^t_{\; t}$.    This rules out the value of $b=\frac32$ that matches with the Schwarzschild metric in isotropic coordinates.   It is seen then that $b<\frac32$.  From Table \ref{tab:1} we see that for $b\leq \frac54$ we have acceptable models for the Solar System.

\begin{table}
\caption{Asymptotic approximations for various values of parameter $b$}
\label{tab:1}
\begin{tabular}{lllllr}
\hline\noalign{\smallskip}
$\; b \; $ & $\; 8\pi \rho(\bar{r}) \; $ & $\; 8 \pi \rho_s(\bar{r}) $ & $ \; e^{2\Phi(\bar{r})}    \; $ & $     \; m(\bar{r}) \; $ & Acceptable?  \\
\noalign{\smallskip}\hline\noalign{\smallskip}
$\; \frac{\, 3\;}{\;2_{\phantom{l}}}$ & $\; 0 $ & $\; \frac{M^2}{2\bar{r}^4}  $ & $ \; 1 - \frac{2M}{\bar{r}} + \frac{M^3}{\bar{r}^3} $ & $\; \; \; M $ & no \\
$\; \frac{\, 5\;}{\; 4_{\phantom{l}}}$ &$\; \frac{M^2}{2\bar{r}_{\phantom{D}}^4} $&$ \; \frac{M^2}{2\bar{r}^4}\Bigl(1-\frac{M}{\bar{r}}\Bigr) $&$  \; 1 - \frac{2M}{\bar{r}} + \frac{3M^3}{2\bar{r}^3} $ & $M\Bigl(1-\frac{M}{4\bar{r}}\Bigr)  $ & yes\\
$\;\; 1    $  &$\; \frac{M^2}{\bar{r}^4} $&$ \; \frac{M^2}{2\bar{r}^4}\Bigl(1-\frac{2M}{\bar{r}}\Bigr) $&$  \; 1 - \frac{2M}{\bar{r}} + \frac{2 M^3}{\bar{r}^3}$ & $M\Bigl(1-\frac{M}{2\bar{r}}\Bigr)   $ & yes \\
$\; \frac{\, 1\;}{\; 2_{\phantom{l}}}$ &$ \; \frac{2M^2}{\bar{r}^4} $&$ \; \frac{M^2}{2\bar{r}^4}\Bigl(1-\frac{4M}{\bar{r}}\Bigr) $&$  \; 1 - \frac{2M}{\bar{r}} + \frac{3M^3}{\bar{r}^3}$ &  $ M\Bigl(1-\frac{M}{\bar{r}}\Bigr) $ & yes \\
$\; \; 0 $ &$ \; \frac{3M^2}{\bar{r}^4} $&$ \; \frac{M^2}{2\bar{r}^4}\Bigl(1-\frac{6M}{\bar{r}}\Bigr) $&$  \; 1 - \frac{2M}{\bar{r}} + \frac{4M^3}{\bar{r}^3}$ &  $ M\Bigl(1-\frac{3M}{2\bar{r}}\Bigr) $ & yes \\
\noalign{\smallskip}\hline
\end{tabular}
\end{table}

\subsection{A specific model with parameter value:  $b=1$}

If we choose $b=1$, we may easily develop a model with exact (not approximate) solutions.  For this model, we equate $e^{2\Lambda}$ to the value $\, 1 + \frac{2M}{r} + \frac{M^2}{r^2}\, $ so that \begin{equation}
  e^{2\Lambda} = \; \biggl( \; 1 \, + \, \frac{M}r \; \biggr)^2
\end{equation}
In order to satisfy (\ref{ePhiiso}) we choose specifically
\begin{equation}
  e^{2\Phi} = \frac{1+\frac{2M}r}{\Bigl(1+\frac{M}r\Bigr)^4}  \approx  1 - \frac{2M}{r} + \frac{2M^2}{r^2}
\end{equation}
From (\ref{Phiprime}) we solve for $\beta(r)$ and thus determine $C^\mu C_\mu$ and hence $8\pi \rho_s$.  The result is
\begin{equation}
  8\pi \rho_s = \; \frac{M^2}{2r^{4}(1+\frac{M}r)^2(1+\frac{2M}r)^2}
\end{equation}
 The resulting metric in line-element form is given by
\begin{equation}
  ds^2 = -\frac{(1+\frac{2M}r)}{(1+\frac{M}r)^4}dt^2 + \Bigl(1+\frac{M}r \Bigr)^2 \biggl( \; dr^2 + r^2 d\theta^2+ r^2 \sin^2\theta d\phi^2 \; \biggr)  \label{isometric}
\end{equation}

From (\ref{r_rbar}) we see that the regular Schwarzschild-type radial coordinate is obtained by the replacement, $\bar{r} = r + M $.
The resulting metric is found to be  (in line element form):
\begin{equation}  ds^2 \, =  - \, (1-\frac{M}{\bar{r}})^3 (1+\frac{M}{\bar{r}}) \, dt^2 + \frac1{(1-\frac{M}{\bar{r}})^2}\, d{\bar{r}}^2 + \,  \bar{r}^2 d\theta^2 + \bar{r}^2\sin^2\theta \, d\phi^2 \label{metric} \end{equation}
and is defined on the interval $\bar{r}> R$, where $R$ is the Schwarzschild radius of the Sun.  We note that $g_{tt} \approx -(\, 1 - \frac{2M}r \,)$.  For this metric and tetrad, the resulting densities and pressures are
\begin{equation}
  C^\mu C_\mu = \frac{M^2}{\bar{r}^4\Bigl(1+\frac{M}{\bar{r}}\Bigr)^2}  \qquad  8 \pi \rho_s =  \frac{M^2}{2\bar{r}^4\Bigl(1+\frac{M}{\bar{r}} \Bigr)^2} \label{rhos2}
\end{equation}
\begin{equation}
  8\pi \rho = \frac{M^2}{\bar{r}^4}  \qquad\qquad \qquad 8\pi \rho_{dm} =  \frac{M^2(1+\frac{4M}{\bar{r}} + \frac{2M^2}{\bar{r}^2})}{2\bar{r}^4(1+\frac{M}{\bar{r}})^2 } \label{totaldensity}
\end{equation}

\begin{equation}
  8\pi p_r = \frac{M^2(1-\frac{3M}{\bar{r}})}{\bar{r}^4 (1+\frac{M}{\bar{r}})} \qquad  8\pi p_T = \; - \frac{M^2(1-\frac{5M}{\bar{r}}-\frac{5M^2}{\bar{r}^2})}{\bar{r}^4(1+\frac{M}{\bar{r}})^2} \label{pressures}
\end{equation}
This matches the standard external metric of Schwarzschild up to the level of accuracy.  The total mass-energy as a function of $\bar{r}$ is
\begin{equation}
  m(\bar{r}) \; = \; M  \, - \, \frac{M^2}{2\bar{r}} \label{mass}
\end{equation}
All solar system results match since the only difference would be due to $F_p$ given by (\ref{e68}) and for this model, we find that $8\pi F_p\approx \frac{2\epsilon M^3}{\bar{r}^6}$, which is negligible.  Thus, the solar system experiments, as reported by Will \cite{Will} , would agree with this metric just as it does for standard general relativity.  The only difference of note is due to the small difference in acceleration due to (\ref{mass}).  The implies that there is an extra outward acceleration at Mercury which is approximately $5.0\times 10^{-10}$ meters per second per second.  As $\bar{r}$ increases this extra outward acceleration decreases and could be interpreted as an extra inward acceleration in the standard theory.  We note that this is slightly less than the Pioneer anomaly value of $8.7 \times 10^{-10}$ meters per second per second.  We claim that the Pioneer anomaly is mostly explained by (\ref{mass}) and partially explained by thermal forces \cite{Turyshev}.

\section{Conclusion.}

The results of this paper show that our theory can reasonably agree with Solar System tests of general relativity.  The value of the parameter $b$ which we use to represent the $\frac{M^2}{r^2}$ term of $g_{rr}$ is seen to be strictly less than $\frac32$, which is the value of $b$ for general relativity.  There is a need for higher precision tests to determine whether $b$ is indeed less than $\frac32$.

\begin{acknowledgements}
The author is deeply thankful for the long-time collaboration with Professor Dave Pandres, Jr., who began this work and who passed away in August 2017.  The author also thanks Peter Musgrave, Denis Pollney and Kayll Lake for the GRTensorII software package which was very helpful.    The author thanks the University of North Georgia  for travel support.
\end{acknowledgements}

\end{document}